\documentclass[a4paper,11pt]{article}
\usepackage{pos}
\usepackage{graphicx}
\usepackage{subcaption}

\newcommand{\nstep}{n_{\mathrm{step}}}

\newcommand{\dd}{\mathrm{d}}
\newcommand{\Pf}{\mathcal{P}_{\mathrm{f}}}
\newcommand{\Pre}{\mathcal{P}_{\mathrm{r}}}

\newcommand{\DD}{{\rm{D}}}
\newcommand{\tmb}[1]{{\mbox{\tiny{#1}}}}
\newcommand{\NambuG}{Nambu-Got\={o}\;}

\title{Stochastic normalizing flows for Effective String Theory}
\author[a]{Michele Caselle}
\author*[a]{Elia Cellini}
\author[a]{Alessandro Nada}

\affiliation[a]{Dipartimento di Fisica,  Universit\'a degli Studi di Torino and INFN, Sezione di Torino, \\
  Via Pietro Giuria 1, I-10125 Turin, Italy}

\emailAdd{elia.cellini@unito.it}

\abstract{Effective String Theory (EST) is a powerful tool used to study confinement in pure gauge theories by modeling the confining flux tube connecting a static quark-anti-quark pair as a thin vibrating string. Recently, flow-based samplers have been applied as an efficient numerical method to study EST regularized on the lattice, opening the route to study observables previously inaccessible to standard analytical methods. Flow-based samplers are a class of algorithms based on Normalizing Flows (NFs), deep generative models recently proposed as a promising alternative to traditional Markov Chain Monte Carlo methods in lattice field theory calculations. By combining NF layers with out-of-equilibrium stochastic updates, we obtain Stochastic Normalizing Flows (SNFs), a scalable class of machine learning algorithms that can be explained in terms of stochastic thermodynamics. In this contribution, we outline EST and SNFs, and report some numerical results for the shape of the flux tube.}

\FullConference{The 41st International Symposium on Lattice Field Theory (LATTICE2024)\\
 28 July - 3 August 2024\\
Liverpool, UK\\}
\begin{document}
\maketitle

\section{Introduction}
Effective String Theory (EST) is a non-perturbative framework that has emerged over the past few decades as a powerful approach for studying confinement in Yang-Mills theory. In this framework, the chromoelectric flux tube connecting a quark-antiquark pair is modeled as a vibrating string~\cite{Nambu:1974zg, Goto:1971ce, Luscher:1980ac, Luscher:1980fr, Polchinski:1991ax}. Specifically, the correlators of two Polyakov loops at distance $R$ are associated with the full partition function of an EST action:
$$\langle P(0) P^{\dagger}(R)\rangle \sim \int \DD X \; e^{-S_\tmb{EST}[X]}.$$

The main tool used in the analysis of EST is zeta function regularization. Numerous results for the interquark potential, derived through the analytical calculation of the partition function, have been successfully compared with high-precision lattice gauge theory calculations~\cite{Aharony:2013ipa,Brandt:2016xsp,Caselle:2021eir,Caristo:2021tbk,Baffigo:2023rin,Caselle:2024zoh}. However, certain scenarios, such as the study of the shape of the flux tube, present challenges that cannot be addressed using analytical methods. An alternative numerical approach to studying EST involves discretizing the EST functional on the lattice and then employing Markov Chain Monte Carlo (MCMC) methods to sample from the corresponding probability distribution. Nevertheless, due to the strong non-linearity of EST actions, standard MCMC methods suffer from critical slowing down when exploring non-trivial EST regimes~\cite{Caselle:2023mvh}.

To mitigate this problem, refs.~\cite{Caselle:2023mvh,Caselle:2024ent} introduced a novel numerical method based on Normalizing Flows (NFs)~\cite{rezende2015variational,Cranmer:2023xbe}. NFs are a class of deep generative models that can sample uncorrelated configurations from a target distribution~\cite{Albergo1}; furthermore, they can also be used to compute observables, such as partition functions, that are inaccessible with traditional MCMC methods~\cite{Nicoli2021,Nicoli:2023qsl}. Despite these advantages, flow-based samplers face challenges in scaling to state-of-the-art applications~\cite{DelDebbio,Abbott:2022zhs,Abbott:2023thq}. These limitations can be mitigated by combining NFs with non-equilibrium simulations based on out-of-equilibrium statistical mechanics~\cite{Caselle:2016wsw,CaselleSU3,Francesconi:2020fgi,Bulgarelli:2023ofi,Bulgarelli:2024onj,Bonanno:2024udh,Bonanno:2024fkn} to obtain Stochastic Normalizing Flows (SNFs)~\cite{wu2020stochastic, Caselle:2022acb}. SNFs have demonstrated excellent scalability in both scalar and gauge theories on the lattice, and are currently among the most promising samplers in lattice field theory~\cite{Caselle:2024ent,Bulgarelli:2024yrz,Bulgarelli:2024cqc,Bulgarelli:2024brv}. In ref.~\cite{Caselle:2024ent}, SNFs could indeed overcome the scaling issues faced in the first work on lattice EST~\cite{Caselle:2023mvh}. In this contribution, we will focus on a new observable in EST (already discussed in~\cite{Caselle:2024ent}): the Binder cumulant $U$ which quantifies the non-Gaussianity of a distribution and can be used to study the shape of the chromoelectric flux tube in pure gauge theory.

In this proceeding, we first briefly review the concepts of EST and SNFs; afterward, we present our numerical results and discuss the implications of these findings for EST.

\section{Effective String Theory}
The most general EST action can be written as:
\begin{align}
    S_\tmb{EST} =  \int_\Sigma d^2\xi \sqrt{g}\bigl[\sigma+\gamma_1 \mathcal{R}+\gamma_2 \mathcal{K}^2+\gamma_3 \mathcal{K}^4\cdots\bigr]
\end{align}
where $g$ is the metric induced on the reference worldsheet surface $\Sigma$, $\sigma$ and $\gamma_i$ are coupling constant, $\mathcal{R}$ represent the Ricci scalar, and $\mathcal{K}$ is the extrinsic curvature. The first term of the EST action corresponds to the well-known \NambuG action. The second term, $\mathcal{R}$, is a topological invariant in two dimensions and can be neglected. The third term, $\mathcal{K}^2$, can be eliminated through suitable redefinitions of the fields. The last term, $\mathcal{K}^4$, has been extensively investigated in recent years~\cite{EliasMiro:2019kyf,Caristo:2021tbk,Baffigo:2023rin,Caselle:2024zoh} to study the interquark potential in lattice gauge theories and is regarded by the EST community as the first non-trivial term beyond the \NambuG EST. The fact that this correction appears at such a large order in the large distance expansion of  the EST is known as "low-energy universality"~\cite{Aharony:2013ipa} and is one of the most important recent results in this context. 
In the following we shall study the corrections induced by this $\mathcal{K}^4$ term to the width and shape of the effective string.
Besides this, for completeness, we shall also study the corrections induced by the $\mathcal{K}^2$ term. Even if in standar ESTs this term can be eliminated, there are situations in which it can nevertheless  play a role. For instance in the model discussed in~\cite{Polchinski:1992ty} (which has been proposed to describe the confining regime of trace-deformed lattice gauge theories~\cite{Unsal:2008ch}) in which the \NambuG term is treated as a perturbation of the $\mathcal{K}^2$ term).

Following these considerations,
in this work, we studied the \NambuG (NG) theory $S_\tmb{NG}$ and the following "Beyond \NambuG" (BNG) actions:
\begin{align}
    S^1_\tmb{BNG} &= S_\tmb{NG}+S_\tmb{$\mathcal{K}^2$} \label{eq:NGK2action}\\
    S^2_\tmb{BNG} &= S_\tmb{NG}+S_\tmb{$\mathcal{K}^4$} \label{eq:NGK4action}.
\end{align}
In order to study these theories numerically, we regularized the actions using the usual conventions; for a target $d=2+1$ target gauge theory, the $D=3$ \NambuG goto action takes the form of:
\begin{equation}
\label{eq:NG}\begin{split}
    S_\tmb{NG}(\phi) &= \sigma \sum_{x \in \Lambda} \biggl(\sqrt{1+(\partial_{\mu}\phi(\tau,\epsilon))^2/\sigma}-1\biggr)  \\
    (\partial_{\mu}\phi(x))^2 &= \bigl(\phi(\tau,\epsilon)-\phi(\tau-1,\epsilon)\bigr)^2+\bigl(\phi(\tau,\epsilon)-\phi(\tau,\epsilon-1)\bigr)^2
    \end{split}
\end{equation}
where $\Lambda$ is a square lattice with size $L\times R$ and index $x=(\tau,\epsilon)$ representing the worldsheet, $\phi(x)$ is a real scalar field describing the transverse degree of freedom of the string. In order to mimic the boundary conditions of Polyakov loop correlators in Lattice Gauge Theories (LGTs), we fix Dirichlet boundary conditions along the $\epsilon$ direction and periodic boundary conditions along the $\tau$ direction as discussed in refs.~\cite{Caselle:2023mvh,Caselle:2024ent}. For the BNG theories, we approximate $\mathcal{K}^2$ at the first order:
\begin{equation}
\label{eq:LK2}
    \mathcal{K}^2 \sim \bigl((\partial_0\partial_0 \phi(x))^2+(\partial_1\partial_1 \phi(x))^2+2(\partial_1\partial_0 \phi(x))^2 \bigr).
\end{equation}
Thus, the discretized BNG terms can be regularized as follows:
\begin{align}
    S_\tmb{$\mathcal{K}^2$}(\phi) &=\gamma_2 \sum_{x \in \Lambda} \mathcal{L}_\tmb{$\mathcal{K}^2$}(\phi(x)) \label{eq:K2} \\
    S_\tmb{$\mathcal{K}^4$}(\phi) &=\gamma_3 \sum_{x \in \Lambda} \bigl(\mathcal{L}_\tmb{$\mathcal{K}^2$}(\phi(x))\bigr)^2 \label{eq:K4}
\end{align}
where:
\begin{equation}
\begin{split}
   \mathcal{L}_\tmb{$\mathcal{K}^2$}(\phi(\tau,\epsilon))&=\bigl(\phi(\tau+1,\epsilon)-2\phi(\tau,\epsilon)+\phi(\tau-1,\epsilon)\bigr)^2+\\
   &+\bigl(\phi(\tau,\epsilon+1)-2\phi(\tau,\epsilon)+\phi(\tau,\epsilon-1)\bigr)^2+\\
   &+\frac{1}{8}\bigl(\phi(\tau+1,\epsilon+1)+\phi(\tau-1,\epsilon-1)-\phi(\tau+1,\epsilon-1)-\phi(\tau-1,\epsilon+1)\bigr)^2
\end{split}
\end{equation}
Due to the Dirichlet boundary conditions, the non-vanishing terms in $\epsilon=R$~\cite{Caselle:2024ent} are:
\begin{equation}
    %\begin{split}
    \mathcal{L}_\tmb{$\mathcal{K}^2$}(\phi(\tau,R))=\bigl(\phi(\tau,\epsilon-2)-2\phi(\tau,\epsilon-1)\bigr)^2+\frac{1}{2}\bigl(\phi(\tau-1,\epsilon-1)-\phi(\tau+1,\epsilon-1)\bigr)^2
    %\end{split}
\end{equation}
\subsection{Observables}
In this contribution, we focus our attention on a new observable defined as a combination of the quartic and quadratic moments of the flux density in the midpoint $\phi(\tau,R/2)$:
\begin{equation}
\label{eq:binder}
    U=1-\frac{\langle \phi^4(\tau,R/2)\rangle_{\tau}}{3\langle \phi^2(\tau,R/2)\rangle_{\tau}^2}.
\end{equation}
Where the expectation values $\langle \dots \rangle_{\tau}$ are computed on the spatial coordinate to $R/2$ averaging over $\tau$ also. The quantity $U$ is generally called the Binder cumulant and, since it is identically zero if $\phi(\tau,R/2)$ follows a Gaussian distribution, it can be used as a probe of the non-Gaussianity of a distribution. In the lattice gauge theory framework, $U$ can be used to study the shape of the flux tube; however, in the EST picture, this observable is inaccessible to standard analytical calculations.

\section{Stochastic Normalizing Flows}
Stochastic Normalizing Flows (SNFs)~\cite{wu2020stochastic,Caselle:2022acb} are a class of deep generative models that combine non-equilibrium simulations~\cite{Caselle:2016wsw} and flow-based sampler~\cite{Cranmer:2023xbe}. SNFs build thermodynamics trajectories between two thermodynamics state $q_0=\exp{(-S_0)}/Z_0$ and $p=\exp{(-S)}/Z$ by means of a composition of parametric diffeomorphisms $g^\theta_n$ and Markov Chain Monte Carlo (MCMC) updates with different transition probabilities $P_{c(n)}$:
$$
   \phi_0 \stackrel{g^\theta_1}{\longrightarrow} \; g^\theta_1(\phi_0) \;
  \stackrel{P_{c(1)}}{\longrightarrow} \; \phi_1 \;
  \stackrel{g^\theta_2}{\longrightarrow} \; g^\theta_2(\phi_1) \;
  \stackrel{P_{c(2)}}{\longrightarrow} \; \phi_2 \;
  \stackrel{g^\theta_3}{\longrightarrow} \; \dots \;
  \stackrel{P_{c(\nstep)}}{\longrightarrow} \; \phi_{\nstep} \equiv \phi
$$
where $c(n)$ define a protocol that interpolates between $q_0$ and $p$.
The forward $\Pf$ probability distribution of each sequence $[\phi_0,\phi_1,\dots,\phi]\equiv \Phi$ can be written as:
\begin{equation}
    \Pf[ \Phi] = \prod_{n=1}^{\nstep}\delta(\phi_{g^\theta_{n}}-g^\theta_n(\phi_{n-1})) P_{c(n)} (\phi_{g^\theta_{n}} \to \phi_n) \, ;
\end{equation}
with $\phi_{g^\theta_{n}}=g^\theta_n(\phi_{n-1})$; similarly, the reverse probability distribution $\Pre$ can be written as:
\begin{equation} 
    \Pre[\Phi] = \prod_{n=1}^{\nstep} P_{c(n)} (\phi_{n} \to \phi_{g^\theta_{n}}) \delta(\phi_{n-1}-g^{-1}_n(\phi_{g_{n}})).
\end{equation}
Thanks to the Crooks' theorem~\cite{Crooks_1999}, the fluctuation relation for this algorithm is~\cite{Vaikuntanathan_2011,Caselle:2022acb,Bulgarelli:2024cqc}:
\begin{equation}
    \frac{q_0(\phi_0)\Pf[\Phi]}{p(\phi)\Pre[\Phi]}=e^{W^\theta_d(\Phi)}
\end{equation}
where $W^\theta_d(\Phi)=W^\theta(\Phi)-\Delta F$ is the dimensionless dissipated work, and the work $W$ can be computed as:
$$W^\theta(\Phi)=S(\phi)-S_0(\phi_0)-Q(\Phi)-\log J_{g^{\theta}}$$
where
$$Q(\Phi)=\sum_{n=0}^{\nstep-1} \left\{ S_{c(n+1)}\left[\phi_{n+1}\right] - S_{c(n+1)}\left[\phi_n\right] \right\}$$
and
$$\ln J_{g^{\theta}}=\sum_{n=1}^{\nstep} \ln J_{g^{\theta}_n}(\phi_{n-1})$$
where $J_{g^{\theta}_n}$ is determinant of the Jacobian of the transformation $g^{\theta}_n$.
Observe that, since the sequences $\Phi$ start from configurations drawn from $p$, computing observables $\mathcal{O}$ over $p$ and $p\Pre$ is equal:
\begin{equation}
     \langle \mathcal{O}(\phi)\rangle_p = \int \dd \phi p(\phi) \mathcal{O}(\phi)=\int \dd \phi \cdots \dd \phi_0 p(\phi)\Pre [\Phi]\mathcal{O}(\phi)=\langle \mathcal{O}(\Phi) \rangle_{\textrm{r}} 
\end{equation}
where the average $\langle \cdots \rangle_{\textrm{r}} $ is computed over all the possible reverse sequences of the SNF. However, we do not have direct access to $\Pre$ but only to $\Pf$ for which the sampled configurations can lie arbitrarily far from equilibrium. Nevertheless, we can generate an equilibrium ensemble using an independent Metropolis-Hasting sampler. In this method, a Markov chain sequence is built by sampling proposals $\phi’$ according to $q_0(\phi’_0)\Pf[\Phi’]$; at the step $j$, $\phi’$ is accepted with probability:
%For each step $j$ of the sequence, a proposal $\phi'$ is generated sampling accordingly to $q_0(\phi'_0)\Pf[\Phi']$ and accepted with probability:
%\begin{equation}
 %   A(\phi^{(j-1)},\phi')= \textrm{min} \biggl(1, \frac{q(\phi_0^{(j-1)})\Pf[\Phi^{(j-1)}]}{p(\phi^{(j-1)})\Pre [\Phi^{(j-1)}]}\frac{q(\phi'_0)\Pf(\Phi')}{p(\phi')\Pre [\Phi']} \biggr)
%\end{equation}
\begin{equation}
    A(\phi^{(j-1)},\phi')= \textrm{min} \biggl(1, \frac{p(\phi')\Pre [\Phi']}{p(\phi^{(j-1)})\Pre [\Phi^{(j-1)}]}\frac{q(\phi_0^{(j-1)})\Pf[\Phi^{(j-1)}]}{q(\phi'_0)\Pf[\Phi']} \biggr)
\end{equation}
Thanks to the SNF's fluctuation relation we can compute the non-trivial term as:
\begin{equation}
    \frac{p(\phi')\Pre [\Phi']}{q(\phi'_0)\Pf[\Phi'] }\frac{q(\phi_0^{(j-1)})\Pf[\Phi^{(j-1)}]}{p(\phi^{(j-1)})\Pre [\Phi^{(j-1)}]} = e^{W(\Phi^{(j-1)})-W(\Phi')}
\end{equation}

This sampling procedure is the non-equilibrium counterpart of the sampler proposed in ref.~\cite{Albergo1} and is referred to as the Non-Equilibrium independent Metropolis-Hastings (NE-iMH) method. Note that, since the proposals are independent, each time a new configuration is accepted, the autocorrelation of the NE-iMH is refreshed. 

SNFs are trained by minimizing the Kullback-Leibler divergence %$\langle W^\theta\rangle_{\textrm{f}}$ (Kullback-Leibler divergence).
$$D_{\textrm{KL}}(q_0\Pf^\theta \| p\Pre^\theta)=\langle W^\theta\rangle_{\textrm{f}}-\Delta F$$
with respect to the parameters $\theta$ of the flow layers $g_n$, where the average $\langle \cdots \rangle_{\textrm{f}} $ is computed over all the possible forward sequences.

\section{Numerical Results}
\begin{figure}[h]
  \centering
  \includegraphics[scale=0.45,keepaspectratio=true]{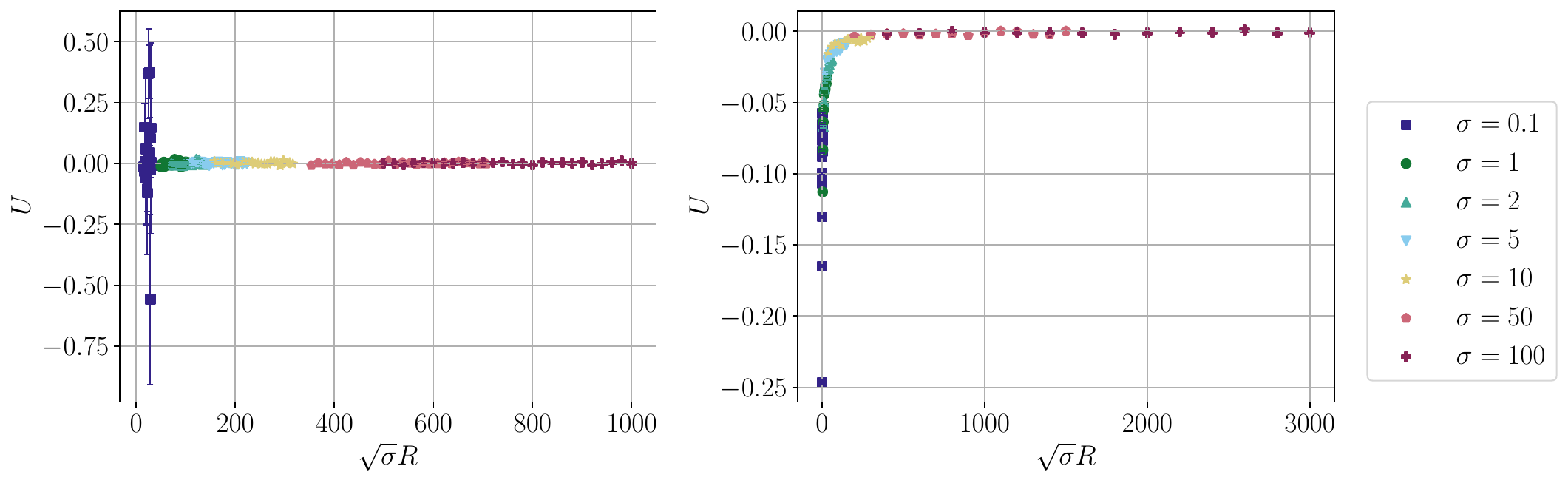}
  \caption{ Binder cumulant $U$ as a function of $\sqrt{\sigma}R$ in the high-temperature limit ($R\gg L=8$, left plot) and in the low-temperature limit ($L=80\gg R$, right plot). Different values of the string tension $\sigma$ are listed in different colors.}
  \label{fig:UNGsigmaLR}
\end{figure}

As mentioned in the introduction, we computed for the first time the Binder cumulant $U$  from eq.~\ref{eq:binder} for EST by sampling configurations using the NE-iMH discussed in the previous section. In fig.~(\ref{fig:UNGsigmaLR}), we present the results for the NG theory from eq.~(\ref{eq:NG}) in both the high and low-temperature regimes. The cumulant vanishes for large values of $\sigma$, indicating that in this regime, we are approaching the Free Boson limit of the \NambuG theory, where the flux tube profile is expected to be Gaussian. Notably, in the low-temperature setup, as the string tension decreases, the cumulant becomes negative. This suggests that the flux tube profile (i.e., the field variable $\phi$) deviates from Gaussian behavior once the Free Boson approximation no longer holds. However, as $R$ increases, this trend weakens progressively. Furthermore, in the high-temperature regime, where $R$ is typically large, $U$ consistently remains zero. This behavior implies that a finite-size effect, driven by Dirichlet boundary conditions, might account for a nonzero $U$~\cite{Caselle:2024ent}.

\begin{figure}[h]
    \centering
    \begin{subfigure}{0.45\textwidth}
        \includegraphics[scale=0.5,keepaspectratio=true]{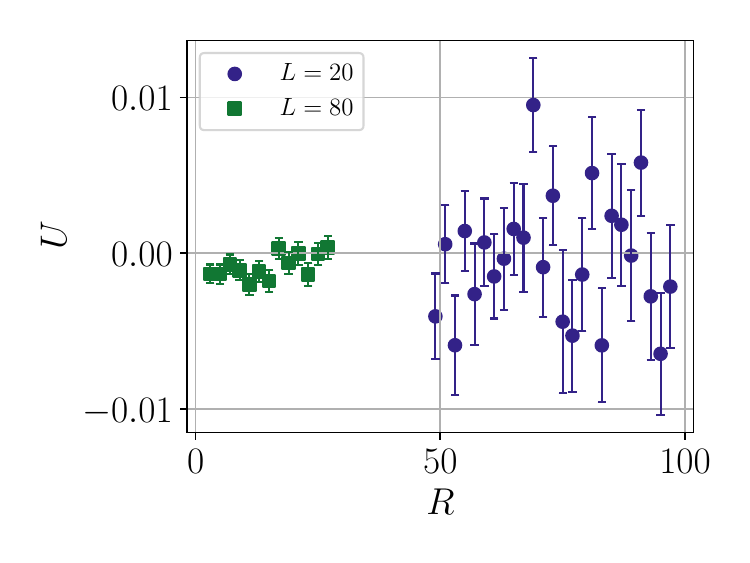}
    \end{subfigure}
    \hfill
    \begin{subfigure}{0.45\textwidth}
        \includegraphics[scale=0.5,keepaspectratio=true]{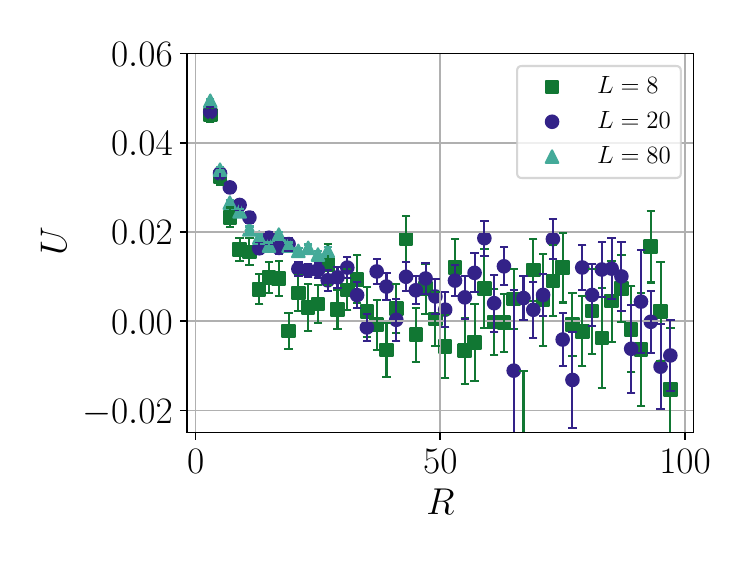}
    \end{subfigure}
    \caption{Binder cumulant $U$ for $S^1_\tmb{BNG}$~(\ref{eq:NGK2action}) at fixed (left plot) and for $S^2_\tmb{BNG}$~(\ref{eq:NGK4action}) at fixed $\gamma_2=0.02$ and $\gamma_3=0.02$. Both theories are studied at $\sigma=100$.}
\label{fig:kbinder}
\end{figure}

Figure~(\ref{fig:kbinder}) displays $U$ for the $S^1_\tmb{BNG}$~(\ref{eq:NGK2action}) (left) and $S^2_\tmb{BNG}$~(\ref{eq:NGK4action}) (right) theories. For the $\mathcal{K}^2$ action (the “rigid” string), no significant deviations from $U=0$ are observed. In contrast, for the $\mathcal{K}^4$ action, a slight deviation from a vanishing Binder cumulant emerges for small values of $R$. Interestingly, this deviation is positive, occurring in the opposite direction to what is seen in the \NambuG case. However, as for the \NambuG theory, the magnitude of this deviation diminishes as $R$ increases, suggesting it may once again be attributed to boundary effects.

We conclude this section with an important observation. In ordinary LGTs, the Binder cumulant of the flux tube is typically negative and significantly larger in magnitude compared to the values reported in this work. Moreover, its dependence on $R$ is distinctly different. This stark contrast indicates that the flux tube’s shape in LGTs behaves fundamentally differently from that in the ESTs studied in this work~\cite{Caselle:2024ent,Caselle:2025vvs}. A quantitative explanation for this behavior lies in the intrinsic width of the confining flux tube.
%In general, we found $U=0$, indicating that the shape profile predicted by EST is Gaussian. For the NG and $S^2_\tmb{BNG}$ theories, however, we observed a small deviation from $0$ when $R$ is small. These deviations are not physically meaningful and are instead attributed to finite size effects~\cite{Caselle:2024ent}.
\section{Conclusion}
In this contribution, based on~\cite{Caselle:2024ent}, we leverage the scalability of SNFs to overcome the scaling issues faced in ref.~\cite{Caselle:2023mvh}. Due to the broad range of observables that SNFs can compute, we calculate, for the first time in EST, the Binder cumulant $U$. Our findings indicate that the deviations from Gaussian behavior are generally very small, with a dependence on $R$ that is consistent with finite size effects, indicating  $U$ effectively zero in the EST framework. In lattice gauge theories, the behavior of the Binder cumulant differs significantly from our EST simulations and is characterized by much larger values; supporting a scenario in which the shape of the flux tube in LGTs cannot be fully captured by EST alone, but instead requires an additional contribution, known as the “intrinsic width”~\cite{Caselle:2025vvs}.

The method used in this contribution could be extended to investigate other models, both in the context of more complex EST and for studying other types of effective theories. In particular, a promising direction for future research is the extension of our model to the $d=3+1$ case to study four-dimensional lattice gauge theories. Another intriguing candidate for further numerical study is the Polchinski-Yang limit of the rigid string~\cite{Polchinski:1992ty}, for which an exact solution for the partition function exists, however, information on the width and shape remains unavailable.

\acknowledgments
We thank Andrea Bulgarelli, Marco Panero, Dario Panfalone and Lorenzo Verzichelli for several insightful discussions. We acknowledge support from the SFT Scientific Initiative of INFN. This work was supported by the Simons Foundation grant 994300 (Simons Collaboration on Confinement and QCD Strings) and partially supported by the European Union - Next Generation EU, Mission 4 Component 1, CUP D53D23002970006, under the Italian PRIN “Progetti di Ricerca di Rilevante Interesse Nazionale – Bando 2022” prot. 2022TJFCYB. 
\bibliographystyle{JHEP}
\bibliography{biblio}

\end{document}